\documentclass
[twocolumn,showpacs,preprintnumbers,amsmath,amssymb,prb,superscriptaddress]{revtex4}
\usepackage{amsfonts}
\usepackage{amsmath}
\usepackage{amssymb}
\usepackage{graphicx}
\usepackage{dcolumn}
\usepackage{mciteplus}
\providecommand{\U}[1]{\protect\rule{.1in}{.1in}}

\begin{document}
\title[KFe$_{2}$Se$_{2}$]{Doping-induced metallicity and coexistence of magnetic subsystems in {K$_{2}
$Fe$_{4+x}$Se$_{5}$}}
\author{Liqin Ke}
\affiliation{Ames Laboratory USDOE, Ames, IA 50011}
\author{Mark van Schilfgaarde}
\affiliation{Department of Physics, King's College London}
\author{Vladimir Antropov}
\affiliation{Ames Laboratory USDOE, Ames, IA 50011}
\keywords{exchange coupling, frustrations, KFe$_{2}$Se$_{2}$, nesting}
\pacs{PACS number}

\begin{abstract}
Electronic structure calculations are used to analyze the electronic and
magnetic properties in {K$_{2}$Fe$_{4+x}$Se$_{5}$}.  Fe atoms can be divided
into two distinct groups.  The $x{=}0$ (parent) compound forms an insulating,
collinear, local moment phase with high N{\'{e}}el temperature.  We show that
large biquadratic exchange coupling and exchange-elastic interactions
stabilize the magnetic order.  For $x{>}0$ the additional Fe atoms fill vacancy
sites.  They form impurity bands for small $x$, which broaden as $x$ increases.
They determine the states at the Fermi level and may be viewed as a magnetic
subsystem separate from the host.  Spin fluctuations are prevalent because
magnetic interactions between the `defect' and the `parent' atoms are
relatively weak, while chemical fluctuations are prevalent for low $x$.
Fluctuations of either type leads to the formation of a weakly metallic state.
The unusual coexistence of the two magnetic subsystems offers a new
perspective as to how superconductivity and strong antiferromagnetism can
coexist.  We argue that spin fluctuations of the impurity subsystem share
common features with the Fe-pnictide superconductors.

\end{abstract}
\eid{identifier}
\date{\today}
\maketitle

\begin{figure}[ptbh]
\includegraphics[scale=0.425]{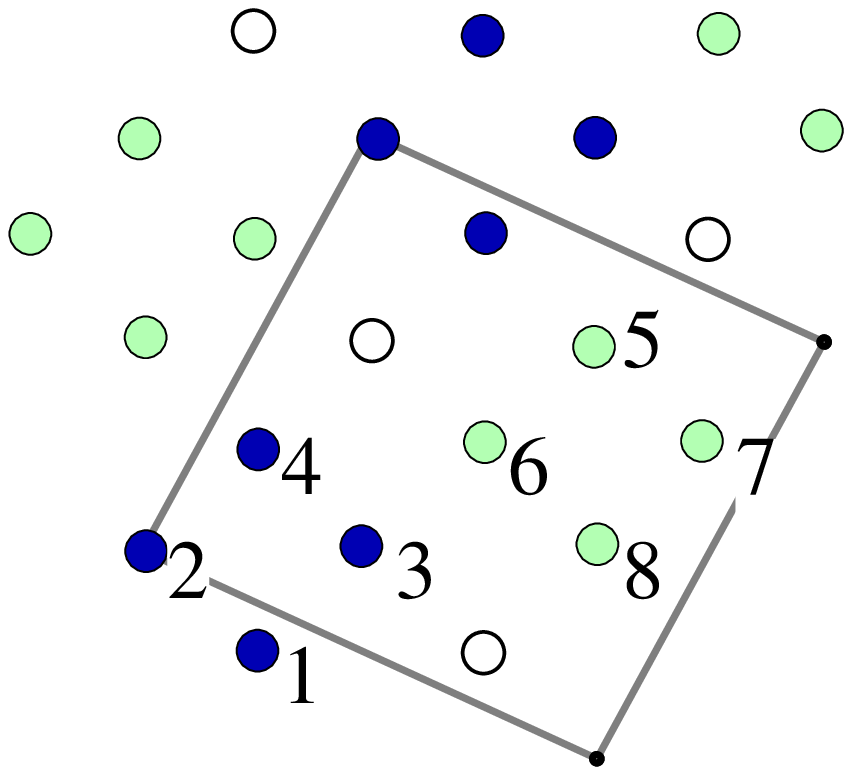} \quad\quad
\includegraphics[scale=0.250]{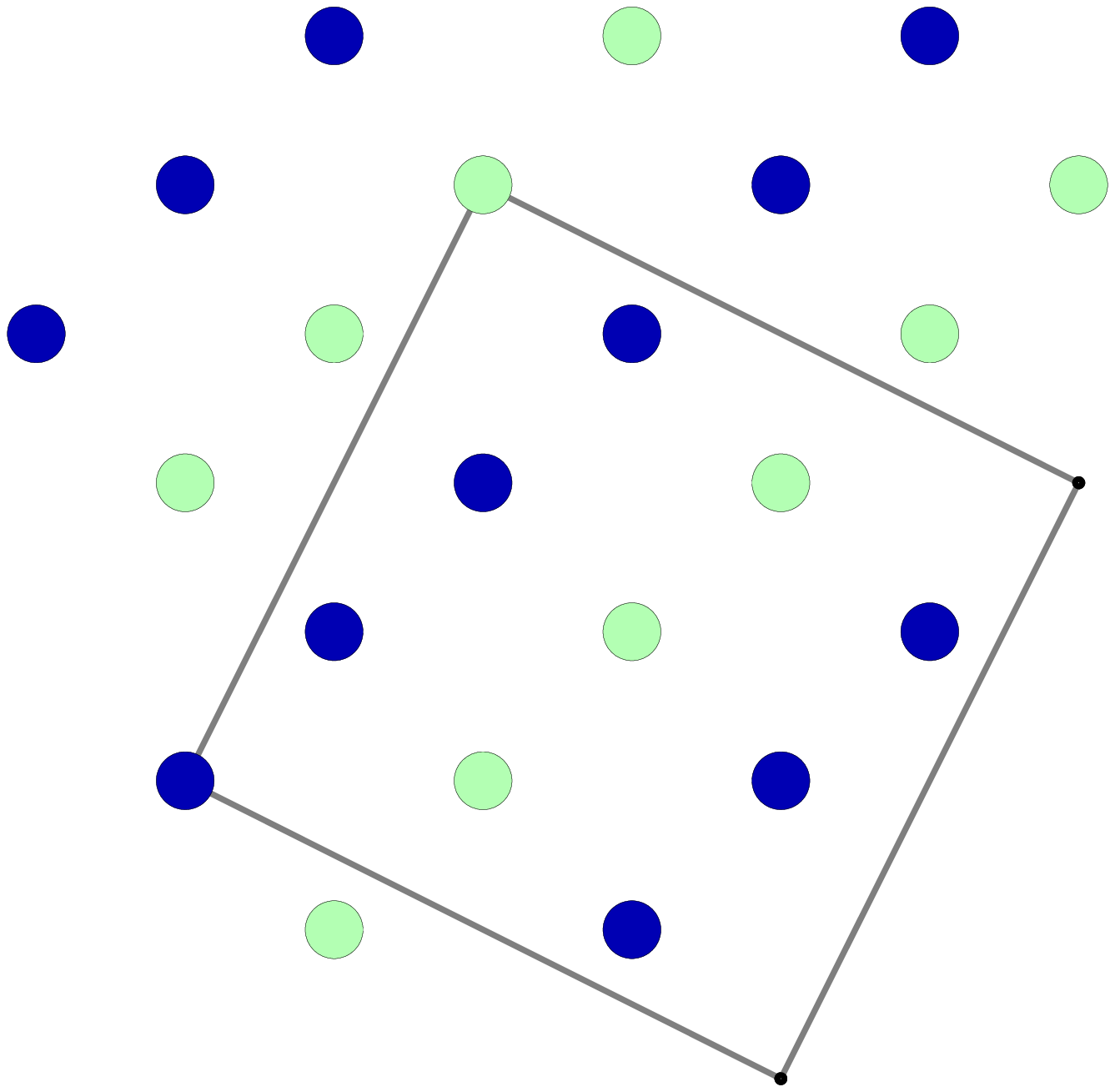}
\caption{Magnetic configuration of {K$_{2}$Fe$_{4+x}$Se$_{5}$} in the block
N{\'e}el structure.  Parallelepiped shows unit cell.  Fe atoms in the \emph{xy}
plane are shown: dark (blue) and light (green) depict spin-up and spin-down Fe
atoms, respectively.  There is a slight distortion of the square (sites 1-4),
elongating (shortening) the diagonal along $x$ ($y$).  Open circles depict 4d
sites that are empty for $x{=}0$, but get populated as $x$ increases.
Relaxation significantly stabilizes the magnetic order.  Right panel shows same
atoms in the striped AFM configuration found in the Fe pnictides, with $x{=}
1$.}
\label{fig:strux}
\end{figure}

The iron selenide superconductors {K$_{2}$Fe$_{4+x}$Se$_{5}$} discovered
recently \cite{SEDISC}, demonstrate superconducting properties similar to
their pnictide counterparts, even though they appear to have very different
magnetic ordering\cite{BAO3} (compare Figs.~\ref{fig:strux}\emph{a},
\ref{fig:strux}\emph{b}) and electronic structure near the Fermi level
$E_{F}$ \cite{ARPES,*qian.prl2011}.  This has a bearing on the associations made between
particular characteristics found in the pnictides, whose properties are
largely invariant from one material to another, to the superconductivity
observed.  Spin interactions are likely to play a key role in mediating
superconductivity; yet their character is not yet well understood
\cite{REVJohnson}.  While biquadratic exchange coupling\cite{BIQNAT} makes
possible a consistent description of magnetism in pnictides which form the
striped antiferromagnetic (AFM) ground state
(Fig.~\ref{fig:strux}\emph{b}), it is not clear what interactions are
responsible for the stabilization of the unusual magnetic structure
observed in selenides \cite{BAO3} (Fig.~\ref{fig:strux}\emph{a}).  In
addition, while weak (and most likely itinerant) AFM ($T_{N}{\sim}
30{\hbox{-}}40$K) and weak superconductivity ($T_{c}{\sim}1{\hbox{-}}10$K)
have been observed to coexistence in the pnictides\cite{REVJohnson}, the
situation is altered in this material.  Magnetic moments and N{\'{e}}el
temperature are much larger, strongly favoring the local moment picture,
and the coexistence of AFM with superconductivity is much more robust
($T_{N}{\sim}640$K and $T_{c}{\sim}30$K)\cite{BAO2,BAO3,LIU}.

In spite of these differences, we will explain how it comes about that
{K$_{2}$Fe$_{4+x}$Se$_{5}$} is metallic, how these two families share some key
features in common, and further how these common characteristics can
potentially provide a framework to explain why these quite distinct material
classes show similar superconductivity, while having quite different kinds of
magnetic interactions.  If such an hypothesis is correct, this work elucidates
what the essential elements are for superconductivity to appear.

The crystalline and magnetic structure of the parent compound K$_{2}$Fe$_{4}
$Se$_{5}$ have been studied in Ref.\cite{CAO,*cao.prl2011}.  Vacancies substitute for 20\%
of the Fe sites in the unusual ordered superstructure shown in
Fig.~\ref{fig:strux}\emph{a}.  The Fe are arranged in squares of four atoms in
a block with the spins parallel.  The blocks themselves are aligned
antiferromagnetically, with two layers of atoms depicted in the Figure.
K$_{2}$Fe$_{4}$Se$_{5}$ has been identified as a magnetic
semiconductor\cite{BAO1,BAO2,BAO3,CAO,*cao.prl2011}, with the Fe local moment observed to
be ${\sim}{3}\mu_{B}$.  It has been argued that superconductivity appears in
this system as a result of vacancy formation\cite{BAO1,BAO2,BAO3,CAO,*cao.prl2011}.

\begin{figure}[ptbh]
\includegraphics[scale=0.5]{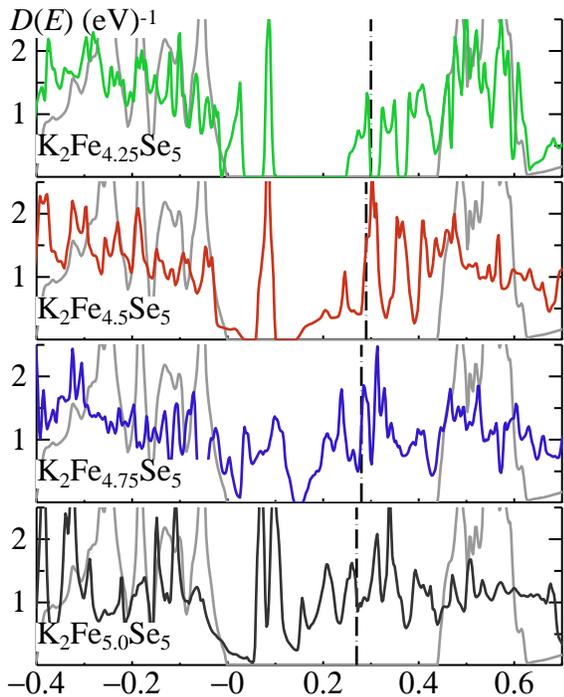}
\caption{Total DOS per Fe atom of {K$_{2}$Fe$_{4+x}$Se$_{5}$} in the collinear
block N{\'e}el structure, for $x{=}0.25$, 0.5, 0.75 and at $x$=1 where all Fe
vacancies are filled.  The parent compound ($x{=}0$) is drawn in light grey,
with a gap of 0.44\,eV and the VBM aligned to 0.  The remaining DOS are
approximately aligned to the parent compound; $E_{F}$ is drawn as a vertical
dot-dashed line for each panel.  As Fe begins to populate the vacancy sites
(top panel), three distinct, very narrow bands appear between 0.25 and 0.4 eV;
$E_{F}$ lies between the first and second.  Another localized state appears at
$E{=}0.08$~eV, and the VBM of the parent compound, splits.  As $x$ increases
from 0.25 to 0.5, the midgap states (especially the state just below $E_{F}$)
broaden but begin to overlap, and when $x{=}0.75$ they merge with the host
bands.}
\label{fig:tdos}
\end{figure}

For the parent system K$_{2}$Fe$_{4+0}$Se$_{5}$, our (LDA) calculation
predicts a magnetic semiconductor in the block N{\'{e}}el structure with a
local moment $M{=}2.9\,{\mu}_{B}$ and a bandgap of 0.44\,eV, confirming the
experiment and the findings of a prior study\cite{CAO,*cao.prl2011}.  This latter work also
concluded that lattice relaxation is necessary to stabilize the block
N{\'{e}}el magnetic structure.  We use the lattice constants of that work; our
LDA implementation is described in Ref.~\cite{PMT}.  We consider how populating
vacancy sites with Fe atoms affects the electronic spectrum and metallicity of
the system, constrained for now to be in a collinear magnetic configuration.
We denote Fe atoms in the parent compound as 16i atoms, and those filling the
vacancy sites as 4d atoms, following customary nomenclature.  Supercells of the
parent structure were generated, and a subset of the vacancy sites populated.
In each case the lattice was relaxed to its minimum-energy configuration,
using the PBE functional. (PBE-relaxed bond lengths, are in better agreement
with experiment LDA ones; the average Fe-Se bond length in K$_{2}$Fe$_{4+0}
$Se$_{5}$ is 2.444\AA , close to the reported value.  But we use PBE only to
relax the structure; the LDA is preferred for magnetic interactions.)
Fig.~\ref{fig:tdos} compares the evolution of total DOS $D(E)$, with $x$
against the parent compound, $x{=}0$.  For $x{=}0.25$, a pair of localized
states form in the gap: the lower band is filled and the upper band is empty.
Both states broaden as $x$ increases to 0.5; still they are nearly separated
so that the system is at best weakly metallic (keeping in mind the LDA tends
to overestimate hybridization and bandwidths).  Magnetic moments range from
2.1-2.3$\mu_{B}$ (Fe on 4d sites) to 2.7-2.9$\mu_{B}$ (16i sites) at
$x{=}0.25$.  Thus in the absence of any fluctuations the impurity band widens,
inducing a transition from semiconducting ($D(E_{F}){=}0$) to weakly metallic
behavior for $x{\gtrsim}0.25$.

\begin{figure}[ptbh]
\includegraphics[scale=0.5]{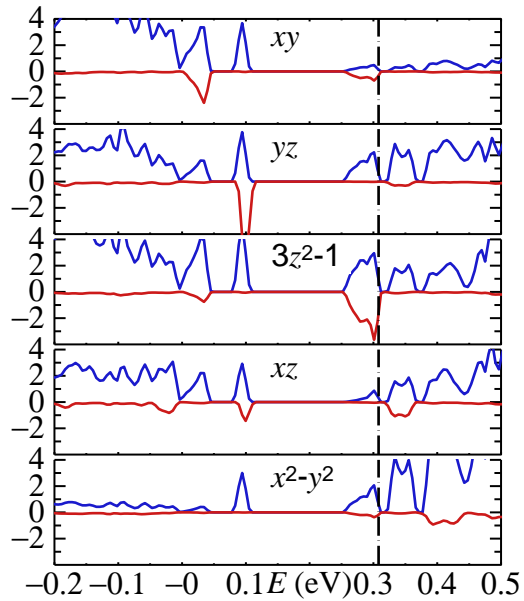}
\caption{Site and \emph{m}-resolved $d$ partial DOS $D(E)$ per Fe atom in
K$_{2}$Fe$_{4.25}$Se$_{5}$. $E_{F}$ is shown as a vertical line: the energy
zero is chosen to align with the VBM of K$_{2}$Fe$_{4}$Se$_{5}$. $D(E)$ above
the zero line are the $m$-resolved DOS summed over the 32 Fe 16i sites,
while DOS below the zero line are summed over the two Fe 4d sites.  The defect
state slightly above $E_{F}$ is confined to the host sites, while he one just
below $E_{F}$, of $3z^{2}{-}1$ character, is centered on the 4d site, with
tails extending to the host.}
\label{fig:mdos}
\end{figure}

In Fig.~\ref{fig:mdos} $D(E)$ is resolved onto $m$ components of $d$ partial
waves at the 4d and 16i Fe sites.  The occupied defect band just below
$E_{F}$ is centered on the Fe 4d $3z^{2}{-}1$ orbital, with tails penetrating
into the host whose cumulative weight approximately matches the head.  The
empty band, on the other hand, is almost completely confined to the host
sites, and is analogous to a ``surface'' resonance.

It is apparent that \emph{chemical disorder} of Fe 4d atoms will lead to
strong fluctuations in the local $D(E_{F})$: in regions where locally $x${$>$
}0.5 defect bands will overlap and the system will be locally metallic.  If the
vacancy occupation is random the local site occupation (concentration) will
follow a binomial distribution, which can be reasonably approximated by a
Gaussian distribution.  Then the probability $P$ of finding a composition
fluctuation in a region containing $N$ ions with amplitude exceeding $\Delta
x_{\mathrm{min}}$ can be written in terms of the standard deviation $\sigma$
as
\begin{align}
P(\Delta x_{\min}; x; N ) = \mathrm{{erfc}}\left(  {{{\Delta x_{\min} }}/
{{\sqrt{2{\sigma^{2}}} }}} \right)  \label{eq:concfluct}
\end{align}
where $\sigma^{2}={{x(1{-}x)}}/{N}$.  The defect wave functions are
sufficiently short ranged that a sphere containing about 8 vacancy sites are
sufficient to stabilize the local DOS.  Partitioning an a compound with
$x$=0.25 average vacancy occupation into overlapping spheres of 8 vacancy
sites each, about 10\% of the spheres would contain a local concentration of
$x{>}0.5$.  Thus from percolation theory we can expect sizable portions of
metallic behavior even at $x{=}0.25$, owing merely to local fluctuations in
the concentration of 4d Fe atoms.

Perhaps more important, there will be strong spin fluctuations on the 4d Fe
sites, as we describe below.  The 16i and 4d Fe atoms contribute to the
electronic structure in approximately independent ways.  The former have large
local moments, are strongly exchange-coupled as described below, and form an
insulating block magnetic N{\'{e}}el structure.  They form relatively wide $d$
bands with a large DOS; however, their contribution to the DOS is shifted away
from $E_{F}$.  Addition of 4d atoms causes defect levels to appear in the
bandgap, which weakly couple to the 16i atoms to form a narrow defect band or
resonance, at $E_{F}$.  Though local moments of these atoms are calculated to
be $2.3\,{\mu}_{B}$ their magnetic coupling to the host is weak.  The reason is
as can be seen in Fig.~\ref{fig:strux}\emph{a}, in a collinear configuration,
spins at the 4d sites must align parallel to one neighboring block and
antiparallel to the other.  The exchange field from the two blocks nearly
cancel, so the net exchange at these sites is weak.  Therefore large,
low-energy spin fluctuations must result.  Such fluctuations can strongly
reduce or even kill the Fe 4d static magnetic moment, making the 4d subsystem
paramagnetic.  This subsystem is thus similar to the entire lattice in other
superconductors such as LiFeAs and FeSe.  In such materials band theory
predicts stable and large moment while experiment clearly shows it is not
present.  In two recent M{\"o}ssbauer studies \cite{MESS,MESSB} `nonmagnetic'
Fe atoms were observed to coexist with the magnetic ones, e.g. in
Ref.~\onlinecite{MESS}, 15\% of Fe atoms in a {K$_{2}$Fe$_{4+0.4}$Se$_{5}$}
sample were reported to be nonmagnetic.

There are a host of recent experiments \cite{BAO3,COEXIST,*zhang.2011arXiv1106.2706Z} demonstrating that
antiferromagnetism and superconductivity coexist.  This has prompted
considerable debate in the literature as to whether the two effects are
present in the same phase, or whether a separate phase coexists that is
responsible for superconductivity.  Several works present experimental evidence
for a second phase; see e.g.  Ref.~\cite{XRAYphase}.  It is suggested that one
phase is insulating in the block magnetic N{\'{e}}el structure that carries
the antiferromagnetism, and the other in a nonmagnetic phase similar to FeSe,
that carries the superconductivity.  There appears to be contradictory
experimental evidence for both the ``one-phase'' and ``two-phase'' scenarios.

The present work cannot rule out either scenario.  But our findings show how it
it is possible that magnetism and superconductivity can coexist in a single
phase.  The 4d Fe atoms generate a rather itinerant channel or subsystem of
states at $E_{F}$ largely decoupled from the 16i magnetic structure.
Fluctuations on the 4d subsystem will be large and the system as a whole
weakly metallic even for small $x$ as a consequence of spin (and chemical)
disorder.  Separation of the 4d and 16i magnetic subsystems is clear, as we
show below, but the electronic states are somewhat mixed and eigenstates at
$E_{F}$ have projections onto both 4d and 16i atoms (Fig.~\ref{fig:mdos}).
Thus superconducting pairing still can originate from electrons of both
subsystems, even while the metallic state is established by the small fraction
of 4d Fe atoms, in a spin (and chemically) disordered configuration.

Next we study the exchange coupling by using a Green's function
linear-response technique, as described in Ref.~\cite{USJIJ}.  Average
values for such linear response $\bar{J}$ are shown in Table 1 for both
ideal and relaxed K$_{2}$Fe$_{4}$Se$_{5}$ structures.  Exchange
interactions more distant than second neighbor are found to be small, and
we present only NN and 2NN parameters.  We distinguish between intrablock
couplings between FM aligned neighbors ($\bar{J}_{1}$, $\bar{J}_{2}$) and
interblock couplings between AFM aligned neighbors ($\bar{J}_{1}^{\prime}
$,$\bar{J}_{2}^{\prime}$).  Relaxation stabilizes the magnetic order, as
can be seen by inspection of the individual $\bar{J}$'s or from the
N{\'{e}}el temperature estimated in the mean field and RPA approximations
(Table 1).

\begin{table}[ptb]
\caption{Average values of first and second NN exchange couplings, $\bar
{J}_{1}$ and $\bar{J}_{2}$, in meV, for K$_{2}$Fe$_{4}$Se$_{5}$.  Interblock
couplings are indicated with a prime, while intrablock couplings are unprimed.
Also shown are the N{\'{e}}el temperature, estimated in the mean-field and
random-phase approximations.}
\begin{tabular}
[c]{|c|r@{\ }|r@{\ }|r@{\ }|r@{\ }|r@{\ }|c|c|} 
          & $M$ & $\bar{J}_{1}$ 
                       & $\bar{J}_{2}$ 
                               & $\bar{J}_{1}^{{\prime}}$ 
                                       & $\bar{J}_{2}^{{\prime}}$ 
                                              & $T_{N}^{MFA}$ & $T_{N}^{RPA}$ \\
\hline
unrelaxed & 2.9 & -5.7 &  17.0 &  25.7 &  6.7 & 682 & ---  \\
relaxed   & 2.9 & -9.7 &  10.2 &  27.3 &  8.5 & 944 & 494  \\ \hline
\end{tabular}
\end{table}

$\bar{J}_{1}$ and $\bar{J}_{1}^{\prime}$ are radically different.  This calls
into question the customary interpretation of these parameters in terms of the
Heisenberg model.  We can extend the Heisenberg hamiltonian to include
biquadratic terms, and assume\cite{BIQNAT} that
\begin{align}
H=\sum_{ij}\bar{J}_{ij}\mathbf{S}_{i}\cdot\mathbf{S}_{j};\quad\bar{J}_{ij}=J_{ij}-2K_{ij}\mathbf{S}_{i}\cdot\mathbf{S}_{j}. \label{eq:biq}
\end{align}
For small angles, where linear response is applicable, $K_{ij}$ can be
eliminated if the intrablock and interblock $J_{ij}$ are permitted to be
different.  It is apparent from the reduced symmetry that ${J}$ need not be
${J}^{\prime}$; moreover for small angles the anisotropic Heisenberg and
biquadratic models are not distinguishable.  They can only be distinguished
at large angles; thus we evaluate the biquadratic term explicitly through
calculations of the total energy $E$ in large-angle noncollinear
configurations.  We accomplish this in practice by rotating the orientation
of four atoms surrounding to one vacancy in a unit cell relative to the
four surrounding the other vacancy (Fig.~\ref{fig:strux}\emph{a}), by a
series of angles $\theta$ ranging between 0 and $\pi$.  Because this
particular rotation preserves all second-neighbor angles, only NN terms in
Eq.~(\ref{eq:biq}) contribute.  A Fourier transform of $E(\theta)$ yields
directly the sum of all pairwise terms in Eq.~(\ref{eq:biq}),
i.e. $J=\sum_{ij}J_{ij}$, $K=\sum_{ij}K_{ij}$ and terms higher order in
$\cos(\theta)$.  Terms beyond the biquadratic are found to be small, so
only $J$ and $K$ need be considered.  As a check, $J+2K$ calculated this
way should match $\bar{J}=\sum_{ij}\bar {J}_{ij}$ calculated by linear
response.  Indeed we find this to be the case: the two calculations agree
to within 3\%.  A large biquadratic coupling $K_{ij}$ of positive sign, on
the same order as $J_{ij}$, is necessary to explain the anisotropy in
$\bar{J}$ and $\bar{J}^{\prime}$.  This system is best characterized by
large positive biquadratic coupling, which initially affects AFM coupling
between blocks and FM coupling inside them.

Lattice relaxation (which depends on $x$) stabilizes FM coupling inside
blocks, as can be seen by its effect on $\bar{J}_{1}$, Table 1.  FM coupling is
further stabilized by local distortions originating from partial filling of
the 4d Fe sites.  Bilinear and biquadratic exchange interactions both change
with relaxation, even while magnetic moment amplitudes are nearly constant.
This effect, which we term the ``exchange-striction'' effect analogous to the
well-known magnetostriction, is in part responsible for stabilizing the
observed magnetic ground state.

The magnetic structure and electronic states at $E_{F}$ in iron selenides
appear to be very different from the Fe pnictides.  Nevertheless they share in
common a large biquadratic coupling, which in each case helps to stabilize the
magnetic ground state structure.  In {K$_{2}$Fe$_{4+x}$Se$_{5}$} this mechanism
is further affected by strong `exchange-striction'.  We showed that two nearly
independent subsystems coexist: a strong AFM phase with large local moments
and high N{\'{e}}el temperature, and a more itinerant phase with fluctuating
moments.  Further, the metallic state (a prerequisite for superconductivity) is
a consequence of fluctuations.  This creates a plausible scenario to explain
the coexistence of superconductivity and strong AFM observed in a single
phase.  In addition, both pnictides and selenides share two key generic
properties: antiferromagnetic interactions create a pseudogap in both cases,
and the DOS at $E_{F}$ is relatively low.  This is also a prerequisite for
spin-mediated superconductivity because a large DOS at $E_{F}$ can be a major
source of incoherent spin scattering, which is usually destructive to
superconductivity originating from spin fluctuations.  The large moments we
find on the 4d subsystem is likely an artifact of density functional theory
--- a mean field approach which does not incorporate fluctuations, and is
expected to be very similar to the well known overestimate of the static
magnetic moment in many iron pnictides systems.  Thus the appearance of such
soft `itinerant' magnetic elements with weakly magnetic or paramagnetic spin
fluctuations can be considered as the only generic feature of both iron
pnictides and chalcogenides, while detailed electronic structure near the
Fermi level, shape of the Fermi surface, magnetic structures and symmetry of
superconducting order parameter seems not universal and not generic.

Work at the Ames Laboratory was supported by US DOE Basic Energy Sciences,
Contract No.  DE-AC02-07CH11358

\bibliography{bbb}

\end{document}